\documentstyle[aps,manuscript,bezier,epsf]{revtex}

\preprint{RCNP-Th00048}

\begin{document}
\draft
\title{\bf Chromomagnetic Catalysis of Color Superconductivity in
a (2+1)-dimensional NJL Model}
\author{{\bf D. Ebert}}
\address{{\it Research Center for Nuclear Physics, Osaka University,
Ibaraki, Osaka 567, Japan}}
\address{{\it and Institut f\"ur Physik, Humboldt-Universit\"at zu
Berlin,
D-10115 Berlin, Germany}}
\author{{\bf K.G. Klimenko}}
\address{{\it Institute for High Energy Physics, 142284
Protvino, Moscow Region, Russia}} 
\author{{\bf H. Toki}}
\address{{\it Research Center for Nuclear Physics, Osaka University,
Ibaraki, Osaka 567, Japan}}

\date{\today}
\maketitle
\begin{abstract}
\baselineskip =0.6cm
The influence of a constant uniform external
chromomagnetic field $H$ on the formation of color superconductivity 
has been investigated.
The consideration was performed
in the framework of a (2+1)-dimensional Nambu--Jona-Lasinio
 model with two different four-fermionic structures responsible
for $<\bar qq>$ and diquark $<qq>$ condensates.
In particular, it was shown that there exists a critical value $H_c$
of the external chromomagnetic field such that at $H>H_c$ a
nonvanishing diquark condensate is dynamically created 
(the so-called chromomagnetic catalysis effect of color
superconductivity).
Moreover, external chromomagnetic fields may in some
cases enhance the diquark condensate of color superconductivity.
\end{abstract} 


\section{Introduction}

During the last two decades a great attention was paid to the
investigation of the QCD ground state at finite temperature and
density
(see e.g. the recent review \cite{smilga} and references therein). 
The main efforts were
directed to the consideration of the quark-gluon plasma -- a new
state of matter which can exist at sufficiently high
temperature.  Besides, it was also realized \cite{love} that at
low (zero) temperature and high baryon density colored
quarks, interacting via gluon exchange, can form Cooper
pairs. Hence, the quark system would pass to the so-called
color superconducting phase in which color symmetry of the
theory is spontaneously broken down.
However, since the corresponding value of the diquark condensate
$<qq>$ was estimated
to be of order 1 MeV, one could not get any observable effects in
this case.

Quite recently it was pointed out \cite{rapp} that due to
instantons there is a nonperturbative mechanism of forming a
condensate
 $<qq>\neq 0$. As a consequence, a rather large observable
value of order 100 MeV for the diquark condensate was predicted
and the color superconductivity (CSC) can possibly be detected in the
future experiments on heavy ion collisions, i. e.  at moderate
baryon density. At present time, there is a rich literature
devoted to this new physical effect: The CSC phenomenon has been
studied in the framework of an instanton model \cite{rapp}, in different
versions of quark models \cite{berges} of the Nambu--Jona-Lasinio (NJL)
 -- type\cite{NJL}, some QCD-like
theories with nonstandard color group and quark multiplets
\cite{kogut} and using lattice and $1/N$ approaches to four-fermion
models \cite{hands}. CSC was also investigated in frameworks of
renormalization group and variational as well as Dyson--Schwinger
equation methods \cite{mir}. In all of the above cited papers
\cite{rapp}-\cite{hands} the nonperturbative feature $<\bar qq>\neq
0$ of
the QCD-vacuum
related to spontaneously broken chiral symmetry was taken into
account. Then,
the phase structure of the theory is the consequence of a competition
between two dynamical order parameters: $<\bar qq>$ and $<qq>$.

It is well-known that the gluonic content of the QCD influences
the properties of the real vacuum, in which there is one more
nonzero nonperturbative condensate $<F^a_{\mu\nu}F^{a\mu\nu}>$
$\equiv$ $<FF>$. Hence, in order to get a more adequate phase structure of the
theory one should consider the competition of three dynamical
parameters $<FF>$, $<\bar qq>$ and $<qq>$. Of course, it is very
hard to solve this task within QCD itself. So, instead of this we shall
incorporate the nonzero gluon condensate $<FF>$ into a simpler NJL-model
consideration of the CSC phenomenon. The NJL model does not contain
dynamical gluons, hence in this case the gluon condensate $<FF>$ is
rather an
external parameter (similar to chemical potential, temperature etc),
than a dynamical one. In the framework of NJL models the condensate
$<FF>\neq
0$ can be realized in terms of an external (background) gauge field
 $A_\mu^a(x)$ \cite{ebert}.

The primary goal of the present paper is just the investigation of
the role which
the gluon condensate will play in the formation of CSC.
In the chosen NJL model approach
we shall, in particular,  consider a
 chromomagnetic gluon condensate, i.e. $<FF>=H^2>0$, with $H$ being
a constant chromomagnetic background field. Let us first comment the 
case of a vanishing diquark condensate.
 One can then imagine that the real vacuum has a color ferromagnet-like
domain structure. Inside each domain the chromomagnetic field $H^a$
is homogeneous, but its direction is varying from one domain to
another in such a way, that space averaging of $H^a$ is equal to
zero. So color and Lorentz invariances are not broken \cite{nielsen}. On the 
other hand, inside the color superconducting phase the SU$_c(3)$ symmetry turns out to be broken spontaneously
to SU$_c(2)$. Using pure symmetry arguments, it is easily
shown
 that the three gluons living in the unbroken SU$_c(2)$ subgroup
stay massless, whereas the remaining five gluons get masses by the
Higgs mechanism.
By analogy with ordinary superconductivity, it is expected that
 external chromomagnetic fields corresponding to massive gluons,
i.e. external chromomagnetic fields of the form
$H^a=(0,0,0,H^4,...,H^8)$,
 should be expelled from the CSC phase (Meissner effect).
Moreover, sufficiently high values of such fields should destroy the
CSC.
 However, our intuition tells us nothing about the action
 of external chromomagnetic fields, which in the color space look
 like
$H^a=(H^1,H^2,H^3,0,...,0)$, on the color superconducting state of
the quark-gluon system.

In the present paper the influence of such external
chromomagnetic fields of the form
$H^a=(H^1,H^2,H^3,0,...,0)$  on the phase structure of the NJL
model is considered. Since we are mainly interested in
clarifying the role of external chromomagnetic fields in the
creation of a color diquark condensate, we put the chemical
potential and temperature equal to zero. For simplicity, all the
considerations are performed in (2+1)-dimensional
space-time.\footnote{Recall that in (2+1)-dimensional
space-time  NJL-type models are renormalizable in the framework
of nonperturbative $1/N_c$ expansion techniques \cite{ros}.
Moreover, these theories are used in order to describe different
planar physical phenomena, including ordinary and high temperature
superconductivity \cite{liu}.}

The paper is organized as follows. In Section II the
NJL model under consideration is presented and its effective
potential at nonzero external chromomagnetic field is obtained
in the one-loop approximation. This quantity contains all the
information about the quark condensates of the theory. In the
following Sections III-V the phase structure of the model is
investigated, firstly for zero background gauge field and then
for nonvanishing vector-potentials of two types 
(Abelian and non-Abelian), 
corresponding
to the same external chromomagnetic field $H$, respectively.  In
Section VI it is shown that there exists a critical value
$H_c$ of the gluon condensate field at which the color diquark
condensate is spontaneously generated (the chromomagnetic
catalysis of CSC). A summary and discussion of the results is given in the
last Section VII. Finally, detailed investigations of global minimum 
points of the effective potential are relegated to the Appendix.

\section{Model Lagrangian and effective potential}

In the present paper the influence of a constant external
chromomagnetic field on the phase structure of an NJL-type of model
with quarks of two flavors and three colors is investigated at zero 
chemical potential and temperature. 
The Lagrangian of the model contains two different
four-fermionic structures responsible for the dynamical appearance of
$<\bar qq>$ as well as $<qq>$ condensates. (Earlier, similar
considerations were done for the simplest NJL-type of models in which
only
a chiral condensate could appear \cite{ebert},\cite{1}-\cite{3}.)
The model under consideration has the following  Lagrangian:
\begin{eqnarray}
\label{eq.1}
L= \bar q\gamma^\mu(i\partial_\mu+eA^a_\mu(x)\frac{\lambda_a}2)
q + \frac{G_1}{6}[(\bar qq)^2+(\bar qi\gamma ^5\vec\tau q)^2]+
\frac{G_2}{3}[i\bar q^C\varepsilon\epsilon^b\gamma^5q][i\bar
q\varepsilon\epsilon^b\gamma^5q_C].
\end{eqnarray}
Here $e$ denotes the gluon coupling
constant, $q_C=C\bar q^T$, $\bar q^C=q^TC$  are charge-conjugated spinors, 
and $C=\gamma^2$ is the charge conjugation
matrix ($T$ denotes the transposition operation). Moreover, 
summation over repeated indices 
$a=1,...,8;$ $b=1,2,3;$ $\mu=0,1,2$ is implied. The quark field
$q\equiv q_{i\alpha}$ is a flavor doublet and color triplet as well as a 
four-component Dirac
spinor, where $i=1,2$; $\alpha=1,2,3$.
(Latin and Greek indices refer to flavor and
color spaces, respectively; spinor indices are
omitted.\footnote{In three dimensions the four-component spinor
representation of the Lorentz group is a reducible one. The
corresponding algebra of $\gamma$-matrices is given for example in
\cite{ros,3}.
Remark also that in our notations $\gamma^5$ is an antisymmetric
matrix, $\gamma^{5T}$=$-\gamma^5$.})
 Furthermore, we use the notations $\lambda_a/2$ for the generators of
the color SU$_c$(3) group as well as $\vec\tau\equiv$
$(\tau^1,\tau^2,\tau^3)$ for Pauli matrices in the flavor space;
and $\varepsilon$ and $\epsilon^b$ are operators in the flavor
and color spaces with matrix elements 
$(\varepsilon)^{ik}\equiv\varepsilon^{ik}$,
$(\epsilon^b)^{\alpha\beta}\equiv\epsilon^{\alpha\beta
b}$, where $\varepsilon^{ik}$ and $\epsilon^{\alpha\beta b}$ are
totally antisymmetric tensors. 
Clearly, the Lagrangian (\ref{eq.1}) is invariant under the color
SU$_c(3)$ and the  chiral SU(2)$_L\times$SU(2)$_R$ groups.

Next, let us at the moment suppose that in (\ref{eq.1}) the $A_\mu^a(x)$
is an arbitrary classical gauge field of the color group SU$_c(3)$.
(The following investigations do not require the explicit inclusion of the 
gauge field part of the Lagrangian).
 Below the detailed structure of $A_\mu^a(x)$
corresponding to a constant chromomagnetic gluon condensate will be
given. Furthermore, in order to demonstrate the main ideas of our approach 
in a way as simple as possible (thereby allowing for other applications in 
planar physics), we shall find it 
convenient to perform the following investigations in the 
(2+1)-dimensional space-time only.\footnote{At nonzero chemical 
potential $\mu$ and
$A_\mu^a(x)=0$, models including diquark channels similar to (1) were 
considered in low
dimensions \cite{hands} and in four-dimensional space-time as well
\cite{berges}, \cite{EB}. In contrast, we consider here the
 case with $A_\mu^a(x)\neq 0$, $\mu=0$.}

Finally, notice that in order to obtain realistic estimates for masses of 
vector/axial-vector mesons and diquarks 
in extended NJL--type of models \cite{EB}, one has to allow for independent 
coupling constants 
$G_1, G_2$, rather than to consider them related by a Fierz transformation
of a current-current interaction via gluon exchange. Clearly, such a  
procedure does not spoil chiral symmetry. 
For the general discussion 
of phase transitions below (see Sections IV-VI) and also 
following the above arguments, we find it 
therefore convenient to treat the coupling constants in 
(\ref{eq.1}) as independent quantities instead of specifying them
by additional model requirements.  

The linearized version of the model (1) with auxiliary bosonic fields
has the following form
\begin{eqnarray}
\label{eq.2}
\tilde L&=&\bar q\gamma^\mu(i
\partial_\mu+eA^a_\mu(x)\frac{\lambda_a}2)
q-\bar q(\sigma + i \gamma^5 \vec\tau\vec\pi)q -
 \frac{3}{2G_1}(\sigma^2+\vec\pi^2)-\frac{3}{G_2}
 \Delta^{\ast b}\Delta^b-\nonumber \\
&-&\Delta^{\ast b}[i
q^TC\varepsilon\epsilon^b\gamma^5q]-\Delta^{b}[i\bar
q\varepsilon\epsilon^b\gamma^5C\bar q^T].
\end{eqnarray}
The Lagrangians (\ref{eq.1}) and (\ref{eq.2}) are equivalent, as can
be seen by using  the
equations of motion for bosonic fields, from which it follows that
\begin{equation}
\label{eq.k200}
\Delta^b\sim i q^TC\varepsilon\epsilon^b\gamma^5q,~~~
\sigma\sim\bar q q,~~~~\vec\pi\sim i\bar q\gamma^5 \vec\tau q.
\end{equation}
Obviously, the $\sigma$ and $\vec\pi$ fields
are color singlets.
Besides, the (bosonic) diquark field $\Delta^b$ is a color
antitriplet and an isospin
singlet under the chiral SU(2)$_L\times$SU(2)$_R$ group. Note further
that  $\sigma, \Delta^b$ are scalars, but 
$\vec\pi$ is a pseudo-scalar field.  Hence, if $\sigma\neq 0$, then
chiral symmetry of the model is spontaneously broken.
$\Delta^b\neq 0$ indicates the dynamical breaking of both the
electromagnetic and the color symmetries of the theory.

In the one-fermion-loop approximation the color and
chirally invariant effective action for the
boson fields is expressed through the path integral over quark
fields:
$$
\exp (i3 S_{\rm eff}(\sigma,\vec\pi,\Delta^b,\Delta^{\ast
b},A_\mu^a))
=
\int [d\bar q] [dq] \exp(i\int\tilde L d^3x),
$$
where
\[
S_{\rm eff}(\sigma,\vec\pi,\Delta^b,\Delta^{\ast b},A_\mu^a)=-
\int d^3x\left [\frac{\sigma^2+\vec\pi^2}{2G_1}+\frac
{\Delta^b\Delta^{\ast
b}}{G_2}\right ]+\tilde S,
\]
and
\begin{equation}
\exp (i3\tilde S) =
\label{eq.3}
\int [d\bar q] [dq] \exp(i\int [\bar q Dq+\bar q\bar M\bar q^T+
q^TMq] d^3x)=\int [d\Psi] \exp(i\int [\Psi^T Z\Psi] d^3x).
\end{equation}
In (\ref{eq.3}) we have used the following notations
\begin{eqnarray}
\label{eq.4}
D=\gamma^\mu(i\partial_\mu+eA^a_\mu(x)\frac{\lambda_a}2)-\sigma-
i\gamma^5\vec\pi\vec\tau,~~
M=-i\Delta^{\ast b}C\varepsilon\epsilon^b\gamma^5, ~~\bar
M=-i\Delta^{b}\varepsilon\epsilon^b\gamma^5C
\end{eqnarray}
and
\begin{equation}
\Psi^T=(q^T,\bar q),~~~~~
Z=\left
(\begin{array}{cc}
M~, & -D^T/2\\
D/2~, & \bar M
\end{array}\right ),
\end{equation}
where $D^T$ denotes the transposed Dirac operator
 (see e.g. the equation (\ref{eq.k3}) below).
 Using the general formula
$$
\det\left
(\begin{array}{cc}
K~, & L\\
\bar L~, & \bar K
\end{array}\right )=\det [-L\bar L+L\bar KL^{-1}K]=\det [-\bar
LL+\bar LK\bar
L^{-1}\bar K],
$$
one can get from (\ref{eq.3}):
\begin{eqnarray}
\label{eq.7}
\exp (i3\tilde S)= {\det}^{1/2}(Z)
={\rm const}\cdot {\det}^{1/2}(D){\det}^{1/2}[D^T+4MD^{-1}\bar M].
\end{eqnarray}

Let us assume that in the ground state of our model
$<\Delta^1>=<\Delta^2>=$$<\vec\pi>=0$ and $<\sigma>$,
$<\Delta^3>\neq 0$.\footnote{Clearly, $<\vec\pi>\neq 0$, would
yield spontaneous breaking of parity. In the theory of
strong interactions parity is, however, a conserved quantum number,
justifying our assumption that $<\vec\pi>=0$. Nevertheless,
 note that at large densities a parity breaking
diquark condensate could appear \cite{par}.}
Obviously, the residual symmetry group of such a vacuum is SU$_c(2)$,
whose generators are the first three generators of initial
SU$_c(3)$. Now suppose that the constant external
chromomagnetic field, simulating the presence of a gluon
condensate $<FF>=H^2$, has the following form
$H^a=(H^1,H^2,H^3,0,...,0)$.
Clearly, due to the residual SU$_c(2)$ invariance of the vacuum, one 
can consider the diquark condensate field $\Delta^a=(0,0,\Delta^3)$, 
putting $H^1=H^2=0$ and $H^3\equiv H$.

Some remarks about the structure of the external condensate 
field $A_\mu^a(x)$, used in (1), are needed. From this moment on, we select
the $A_\mu^a(x)$ in such a form that the only nonvanishing components of
the corresponding field
strength tensor $F^a_{\mu\nu}$ are $F^3_{12}=-F^3_{21}=H$=const.
It is well-known that in
non-Abelian gauge theories a given
external chromomagnetic field does not fix the kind of the
corresponding
vector-potential uniquely. In other words, there can exist several
physically
(gauge) nonequivalent vector-potentials, which produce the same
chromomagnetic field \cite{brown}. For example, in the case under
consideration, i.e. in three dimensions, the above homogeneous
chromomagnetic field
can be generated  by two qualitatively different vector-potentials:
\begin{equation}
\label{eq.k2}
A_\mu^1(x)=(0,\sqrt{H/e},0),~~A_\mu^2(x)=(0,0,\sqrt{H/e})
 ,~~A_\mu^a(x)=0~~(a=3,...,8),
\end{equation}
or
\begin{equation}
\label{eq.k1}
A_\mu^a(x)=H\delta_{\mu 2}x^1\delta^{a3}.
\end{equation}
The second of these vector-potentials defines the
 well-known Matinyan-Savvidy
model of the gluon condensate in QCD \cite{mat}.
In the following we shall denote expressions (\ref{eq.k2}) and
(\ref{eq.k1}) as vector-potentials I and II,
correspondingly.\footnote{Having background fields $A_\mu^a(x)$, 
given by (\ref{eq.k2}) or (\ref{eq.k1}),
one can form three matrix fields
$A_\mu(x)=A_\mu^a(x)\frac{\lambda_{a}}{2}$,
($a=1,2,3$).
Now it is easy to see that for vector-potential I (II) the
corresponding fields $A_\mu(x)$ do not commute (commute) between
themselves. Due to this reason, the vector-potential I (II) is called
sometimes non-Abelian (Abelian) vector-potential.}

There exists an attractive picture of a domain structure of the physical 
vacuum of QCD which assumes that the space
is splitted into an infinite number of macroscopic domains 
each of which containing a homogeneous chromomagnetic background field
generating a nonzero gluon condensate $<FF>\neq 0$.  \cite{nielsen}. 
Averaging over all domains results in 
a zero background chromomagnetic field, hence color as well as
Lorentz symmetries are not broken. (Strictly
speaking, our following calculations refer to some given 
macroscopic domain. The obtained results turn out to depend on
color and Lorentz invariant quantities only,
and are independent of the concrete domain.)
Note also that it was pointed out in \cite{cabo}
that, at high temperature, Abelian-like vector-potentials of the
form (\ref{eq.k1}) may serve as a reasonable approximation to the 
true vacuum of the theory. However, at
low temperature, the background gauge field may be essentially
non-Abelian, having the form (\ref{eq.k2}).

 In order to find nonvanishing condensates $<\sigma>$ and
$<\Delta^3>$, we should calculate the effective potential whose
global minimum point provides us with these quantities.
Suppose that (apart from the external vector-potential $A_\mu^a(x)$
(\ref{eq.k1})) all boson fields in $S_{\rm eff}$ do
not depend on
space-time. In this case, by definition, $S_{\rm eff}=-V_{\rm
eff}\int d^3x$, where
\begin{eqnarray}
\label{eq.8}
V_{\rm eff}&=&\frac{\sigma^2+\vec\pi^2}{2G_1}+\frac
{\Delta^b\Delta^{\ast
b}}
{G_2}+\tilde V,\nonumber\\
\tilde V&=&\frac i{6v}\ln [\det(D)\det[D^T+4MD^{-1}\bar
M]],~~v=\int
d^3x.
\end{eqnarray}
Due to our assumptions on the vacuum structure, in formulae
(\ref{eq.8}) we should put $\Delta^{1,2}\equiv 0$ as well as
$\vec\pi=0$. Then, taking into account the relations
\[
\gamma^{\mu T}C=-C\gamma^\mu,~~ (\mu=0,1,2);~~
\lambda^{aT}\epsilon^3=-\epsilon^3\lambda^a,~~ (a=1,2,3),
\]
we have for the operator $D$ (cf.(\ref{eq.4})) with vector-potentials
(\ref{eq.k2}),(\ref{eq.k1}) the following identity
\begin{equation}
D^TC\epsilon^3\equiv [\gamma^{\mu
T}(-i\partial_\mu+eA^a_\mu(x)\lambda^T_a/2)
-\sigma]C\epsilon^3=C\epsilon^3D.
\label{eq.k3}
\end{equation}
Now, using this formula as well as the relations $\det D=\det
(\gamma^5D^T\gamma^5)$, $\det AB=\det A\det B$,
$(\varepsilon\varepsilon)_{ij}=-\delta_{ij}$,$~(\epsilon^3\epsilon^3)
_{\alpha\beta}=-\delta_{\alpha\beta}+
\delta_{\alpha 3}\delta_{\beta 3}$ in the expression
(\ref{eq.8}),
one can obtain after some evident transformations
(recall that $i,j=1,2$ and $\alpha,\beta=1,2,3$):
\begin{eqnarray}
\label{eq.10}
\tilde V&=&\frac i{6v}\ln
\det[\gamma^5D^T\gamma^5D^T+4\Delta\Delta^\ast\varepsilon\varepsilon
\epsilon^3\epsilon^3]\nonumber\\
&=&\frac i{6v}\ln\det[D\gamma^{5T}D\gamma^{5T}+
4\Delta\Delta^\ast\delta_{ij}(\delta_{\alpha\beta}-\delta_{\alpha
3}\delta_{\beta 3})],
\end{eqnarray}
where $\Delta\equiv\Delta^3$.
Remark, that the first term under the $\det$-symbol in
(\ref{eq.10}) is a diagonal operator in the flavor space.  One
can easily see that the second term there is also a diagonal
operator, but this time in the flavor, color, spinor as well as
coordinate spaces.

\section{Phase structure at zero external field}

If the external field vanishes, we have the evident relation
$D\gamma^5D\gamma^5$=$(\sigma^2+\partial^2)\delta_{ij}\delta_{\alpha
\beta}$.
Taking
into account this formula as well as the usual one, $\det
O=\exp({\rm tr}\ln O)$, the determinant in (\ref{eq.10}) can be
calculated straightforwardly leading to the
following one-loop expression for the effective potential
of the initial model (\ref{eq.8})
($V_{\rm eff}\equiv V_0$):
\begin{eqnarray}
\label{eq.11}
V_0(\sigma,\Delta,\Delta^\ast)&=&\frac{\sigma^2}{2G_1}+\frac
{\Delta\Delta^{\ast}}{G_2}
+\frac{8i}{3}\int
\frac{d^3k}{(2\pi)^3}\ln(\sigma^2+4|\Delta|^2-k^2)\nonumber\\
&+&\frac{4i}{3}\int \frac{d^3k}{(2\pi)^3}\ln(\sigma^2-k^2),
\end{eqnarray}
where we have used the momentum space representation. This
expression has ultraviolet divergences. Hence, we need to
regularize it by introducing in (\ref{eq.11}) Euclidean metric
($k^0\to ik^0$) and cutting off the range of integration
($k^2\leq\Lambda^2$). 
Next, by performing the integration and introducing renormalized coupling 
constants $g$ and $f$, 
defined by
\begin{equation}
\label{eq.12}
\frac{1}{G_1}-\frac{4\Lambda}{\pi^2}
\equiv\frac{1}{g},~~~~
\frac{1}{G_2}-\frac{16\Lambda}{3\pi^2}
\equiv\frac{1}{f},
\end{equation}
we can express 
the effective potential in terms of ultraviolet-finite
quantities 

\begin{eqnarray}
\label{eq.13}
V_0(\sigma,\Delta,\Delta^\ast)=\frac{\sigma^2}{2g}+\frac
{\Delta\Delta^{\ast}}
{f}+\frac
{4}{9\pi}(\sigma^2+4|\Delta|^2)^{3/2}+\frac{2}{9\pi}|\sigma|^3.
\end{eqnarray}

We shall now search for the global minimum point of the potential
(\ref{eq.13}). Before doing this, let us introduce a set of
more convenient parameters
\begin{eqnarray}
\label{eq.15}
\phi=2|\Delta|,~~A=3\pi/g,~~B=3\pi/(2f),
\end{eqnarray}
in terms of which the effective potential
(\ref{eq.13}) is given by
\begin{eqnarray}
\label{eq.16}
3\pi V_0(\sigma,\phi)=\frac{A\sigma^2}{2}+\frac {B\phi^2}
{2}+\frac
{4}{3}(\sigma^2+\phi^2)^{3/2}+\frac{2}{3}|\sigma|^3.
\end{eqnarray}
By symmetry reasons, it is sufficient to study the function
(\ref{eq.16}) in the region $\{\sigma\geq 0, \phi\geq 0\}$,
where we have the following stationarity equations
\begin{eqnarray}
\label{eq.17}
\frac{\partial
V_0}{\partial\sigma}=0=\sigma\{A+4\sqrt{\sigma^2+\phi^2}+2\sigma\},
\nonumber\\
\frac{\partial
V_0}{\partial\phi}=0=\phi\{B+4\sqrt{\sigma^2+\phi^2}\}.
\end{eqnarray}
In order to find the global minimum point of the effective
potential (\ref{eq.16}) one should search for all the solutions
of the stationarity equations (\ref{eq.17}) and then find among them
the single one, where the effective potential takes an absolute
minimum value.
Omitting the detailed calculations, we present at once the
results in terms of the phase portrait shown in Fig. 1.

This figure shows the plane of parameters $(A,B)$,
divided into four domains
corresponding to the four possible phases of the model.
In domain I the point $\sigma,\phi=0$ is
the absolute minimum point of $V_0$, in II it is at values $\phi
=0,\sigma=-A/6$, in the region III the global minimum is located at
the point
$\sigma=0,\phi=-B/4$.  Finally, in region IV the global minimum lies
in the point $\sigma=(B-A)/2,\phi=\sqrt{B^2/16-(B-A)^2/4}$.

Recall, that the coordinates of the global minimum point of the
effective potential are the vacuum expectation values
$<\sigma>,<\phi>$ of the fields $\sigma,\phi$. Lagrangian
(\ref{eq.2})
provides us with the equations of motion for $\sigma,\Delta$ from
which one can easily get the following relations $<\sigma>\sim
<\bar qq>$, $<\phi>\sim <qq>$, i.e. the global minimum point of the
effective potential gives us an information about chiral and
Cooper-type diquark condensates of the model. Hence, if the
parameters
$(A,B)$ belong to the region I, we have the symmetric phase,
because of $<\sigma>=<\bar qq>=0$, $<\phi>=<qq>=0$ in this case.
The phase with spontaneously broken chiral symmetry is situated
in region II, since here the chiral condensate $<\bar qq>\not=0$.
In region III the diquark condensate $<qq>$ is nonzero,
 thus indicating the presence of the color superconductivity phase.
Clearly, in this case the electromagnetic as well as color symmetries
of the
model are spontaneously broken. Finally, region IV
corresponds to the mixed phase of the theory, where chiral,
electromagnetic and color symmetries are spontaneously broken
down (here $<\bar qq>,<qq>\not=0$).

\section{Phase structure of the model for vector-potential I}

Let us next study the influence of the external chromomagnetic
field with constant vector-potential (\ref{eq.k2}) on the phase
structure of the model (1). 
By using the momentum space representation for the operator 
under the $\det$-symbol in (\ref{eq.10}), we
obtain instead of the differential operator an algebraic
(24$\times$24) matrix which has three different eigenvalues
$E_i(p)$ (i=1,2,3)
\begin{equation}
E_{1,2}(p) = \bar p^2-p_0^2 +\sigma^2+4|\Delta|^2+ \frac{eH}{2}\pm
\frac{1}{2}\sqrt{(eH)^2 + 4eH\bar p^2},~
E_3(p) = \bar p^2-p_0^2 +\sigma^2,
\label{eq.18}
\end{equation}
each of them having 
an eight-fold degeneracy,
and $\bar p^2 = p^2_1 + p^2_2$. Taking into account this fact, one
can easily obtain from (\ref{eq.8}) and (\ref{eq.10}) the
following expression for the effective potential of the model in
the presence of an external vector-potential (\ref{eq.k2}) ($V_{\rm
eff}\equiv V_{H_1}$):
\begin{eqnarray}
\label{eq.19}
V_{H_1}(\sigma,\Delta,\Delta^\ast)=\frac{\sigma^2}{2G_1}+
\frac{\Delta
\Delta^{\ast}}{G_2}+\frac{4i}{3}\sum_{i=1}^3\int
\frac{d^3p}{(2\pi)^3}\ln~E_i(p),
\end{eqnarray}
Integrating here first over $p_0$ and then over the
space of $p_{1,2}$ variables and employing a suitable ultraviolet
cutoff,
one can get after adopting a renormalization
procedure (similar calculations were performed in \cite{1} for the
model (1) in the case with $G_2=0$):
\begin{eqnarray}
&&3\pi V_{H_1}(\sigma,\phi) =
\frac{A\sigma^2}{2}+\frac{B\phi^2}2
+\frac{2|\sigma|^3}{3}
+\frac 23\left [(\sigma^2+\phi^2)^{3/2}+(\sigma^2+\phi^2+eH)^{3/2}
 - \right.\nonumber  \\
&-&\frac{3eH}{4}(\sigma^2+\phi^2+eH)^{1/2}-
 \left.\frac 34(\sigma^2+\phi^2)\sqrt{eH} \ln
~\frac {\sqrt{eH} +
\sqrt{\sigma^2+\phi^2+eH}}{\sqrt{\sigma^2+\phi^2}}\right ],
\label{eq.20}
\end{eqnarray}
where we have used the notations introduced in formula
(\ref{eq.15}). Instead of the dimensional quantity (\ref{eq.20}), let
us consider the dimensionless function $V_1(x,y)\equiv 3\pi
V_{H_1}(\sigma,\phi)/(eH)^{3/2}$, where
$x=\sigma/\sqrt{eH}$,  $y=\phi/\sqrt{eH}$. Evidently, we have
\begin{eqnarray}
V_1(x,y)&=&\frac{\tilde Ax^2}2+\frac{\tilde
By^2}2+\frac23|x|^3+\frac23\left
[(x^2+y^2)^{3/2}+
(1+x^2+y^2)^{3/2}\right.-\nonumber\\
&-&\left.\frac34\sqrt{1+x^2+y^2}-\frac34(x^2+y^2)\ln\left
[\frac{1+\sqrt{1+x^2+y^2}}{\sqrt{x^2+y^2}}\right ]\right ],
\label{eq.200}
\end{eqnarray}
where $\tilde A=A/\sqrt{eH}$, $\tilde B=B/\sqrt{eH}$.

\subsection*{Phase structure in terms of $\tilde A$
and $\tilde B$}

In contrast to the potential (\ref{eq.20}), which has three
dimensional parameters $A,B,eH$, the function (\ref{eq.200}) depends
only on two parameters $\tilde A,\tilde B$. So, first of all we shall
study the global minimum dependence of the potential $V_1$ on
the parameters $\tilde A,\tilde B$. Doing this, we get the
possibility
to discuss the phase structure of the model in the presence of
the external vector-potential (\ref{eq.k2}). Since $V_1$ is
symmetric  under transformations $x\to -x$ or $y\to -y$, it is
sufficient to look for its global minimum only in the region
$x,y\geq 0$. The stationarity equations then take the form
\begin{equation}
\frac{\partial V_1}{\partial x}\equiv x\{~\tilde
A+2x+F(z)~|_{z=\sqrt{x^2+y^2}}~\}=0~~,
~~\frac{\partial V_1}{\partial y}\equiv y\{~\tilde
B+F(z)~|_{z=\sqrt{x^2+y^2}}~\}=0,
\label{eq.21}
\end{equation}
where
\begin{equation}
F(z)=2z+2\sqrt{1+z^2}-\ln[~(1 + \sqrt{1+z^2})/z~].
\label{eq.22}
\end{equation}
We should find all the solutions of the equations (\ref{eq.21})
in the region $x,y\geq 0$ and then select that one, where the
function $V_1(x,y)$ takes its global minimum. Omitting here
calculational details, we directly quote the results in the form of
the phase portrait. The detailed investigation
is given in Appendix A.

In  Fig. 2 the
phase portrait of the model (1) in the presence of a nonzero
vector-potential of the type I is presented in terms of $\tilde A$
and $\tilde B$. Here one can see three regions. Above the line
$\tilde l_1$
there is
the Phase III of the model, which corresponds to the $V_1(x,y)$
global minimum point of the form $(0,y_0(\tilde B))$.
(The properties of the function $y_0(\tilde B)$ as well as the
functions
$x_0(\tilde A)$,
$x_1(\tilde A,\tilde B)$, $y_1(\tilde A,\tilde B)$ considered below
are given 
in Appendix A.)
In this case
$<\sigma>=0$, $<\phi>\neq 0$. Below the curve $\tilde l_2$ the global
minimum
of the effective potential has the form $(x_0(\tilde A),0)$. Thus, in
this region the Phase II is located, since for such values of
$\tilde A, \tilde B$ the model has a vacuum with $<\sigma>\neq 0$,
$<\phi>=0$.  Finally, inside the $\Omega$-domain there is a mixed
Phase IV, since here the global minimum point
 $(x_1(\tilde A,\tilde B),y_1(\tilde A,\tilde B))$
corresponds to the vacuum with  $<\bar q q>\neq 0$ and $<qq>\neq 0$.
 It is necessary to emphasize that in the presence of such kind of
external vector-potentials the Phase I is absent at all
(obviously, only when $A_\mu^a(x)=0$, this phase can be realized in the
model).

\section{Phase structure of the model for vector-potential II}

Next, let us study the influence of a nonvanishing external
chromomagnetic
field with vector-potential (\ref{eq.k1}) on the phase
structure of the model (1). In this case, after some calculations, the
operator, which
stands under the $\det$-symbol in (\ref{eq.10}), can be transformed
to the following expression in the color space:
\begin{eqnarray}
\label{eq.28}
D\gamma^{5}D\gamma^{5}+
4\Delta\Delta^\ast\delta_{ij}(\delta_{\alpha\beta}-\delta_{\alpha
3}\delta_{\beta 3})=
\mbox{diag}(\delta_{ij}D_+~,~\delta_{ij}D_-~,~\delta_{ij}(\sigma^2+
\partial^2)),
\end{eqnarray}
where
\begin{eqnarray}
\label{eq.29}
D_\pm=\sigma^2+4|\Delta|^2-\Pi_\mu^\pm\Pi^{\pm\mu}\mp\frac
{ie}4\gamma^\mu\gamma^\nu\bar
F_{\mu\nu},~\Pi_\mu^\pm=i\partial_\mu\pm e\bar A_\mu(x)/2,~\bar
A_\mu=H\delta_{\mu 2}x_1.
\end{eqnarray}
Remark, that in (\ref{eq.28}), (\ref{eq.29}) $D_\pm$ are operators
in the coordinate and spinor spaces, only. The same is true for the 
expression
$(\sigma^2+\partial^2)$, which is the unit operator in the spinor
space.
Taking into account  (\ref{eq.28}), one can easily find for the
potential $\tilde V$ (\ref{eq.10}) the expression:
\begin{eqnarray}
\label{eq.30}
\tilde V=\frac{i}{3v}{\rm tr}\ln D_++\frac{i}{3v}{\rm tr}\ln D_-
+\frac{i}{3v}{\rm tr}\ln (\sigma^2+\partial^2)
\end{eqnarray}
(the trace prescription in (\ref{eq.30}) is taken over coordinate
as well as spinor spaces). The last term in this formula was
calculated in Section III of the present paper. Concerning the
first two terms in (\ref{eq.30}), we should point out, that
$D_\pm=\tilde D_\pm\gamma^5\tilde D_\pm\gamma^5$, where $\tilde D_\pm
=i\gamma^\mu\partial_\mu\pm e\gamma^\mu\bar A_\mu(x)/2+M$ are
formally
the Dirac operators for fermi-particles with electric charges
$\pm e/2$ and effective mass $M=\sqrt{\sigma^2+4|\Delta|^2}$ in the
presence of
a constant external magnetic field $H$. Similar expressions were
calculated in a lot of papers (see, e.g. \cite{2}, from which it
follows that the first term in (\ref{eq.30}) equals to the second
one). So, we omit details of ${\rm tr}\ln D_\pm$ calculations
and present at once the corresponding effective potential of the
model:
\begin{eqnarray}
\label{eq.31}
V_{H_2}(\sigma,\Delta,\Delta^*)=\frac{\sigma^2}{2G_1}+\frac
{\Delta\Delta^{\ast}}{G_2}&+&\frac{eH}{6\pi^{3/2}}
\int_{0}^{\infty} \frac{ds}{s^{3/2}} \exp
(-s(\sigma^2+4|\Delta|^2))~\coth(eHs/2)\nonumber \\
&+&\frac{4i}{3}\int \frac{d^3k}{(2\pi)^3}\ln(\sigma^2-k^2).
\end{eqnarray}
In this formula $eH$ has a positive value. Both integrals in
(\ref{eq.31}) are ultra-violet divergent ones. To renormalize the
first integral
one can act in the same manner, as it was done in \cite{3}
with the effective potential of the three-dimensional Gross-Neveu
model
in the presence of an external magnetic field. The second integral in
(\ref{eq.31}) was already renormalized in Section III. Hence, the
finite expression for the effective potential of the model (1)
in an external chromomagnetic field of type II looks like:
\begin{eqnarray}
\label{eq.32}
V_{H_2}(\sigma,\Delta,\Delta^*)&=&V_0(\sigma,\Delta,\Delta^*)
+\nonumber \\
&+&\frac{eH}{6\pi^{3/2}}
\int_{0}^{\infty} \frac{ds}{s^{3/2}} \exp (-s(\sigma^2+4|\Delta|^2))
\left [\coth\left (\frac{eHs}2\right )-\frac 2{eHs}\right ],
\end{eqnarray}
where $V_0(\sigma,\Delta,\Delta^*)$ denotes the effective potential
at
$H=0$
(see (\ref{eq.13})). Integrating in this formula with the help
of integral tables \cite{prud}, one can get the following more
compact
expression for the effective potential
\begin{eqnarray}
\label{eq.33}
V_{H_2}(\sigma,\Delta,\Delta^*)&=&\frac{\sigma^2}{2g}+\frac
{\Delta\Delta^{\ast}}
{f}+\frac{2}{9\pi}|\sigma|^3+\nonumber \\
&+&\frac{eH\sqrt{\sigma^2+4|\Delta|^2}}{3\pi}-
\frac{2(eH)^{3/2}}{3\pi}\zeta\left
(-\frac12,\frac{\sigma^2+4|\Delta|^2}{eH}\right),
\end{eqnarray}
where $\zeta(\nu,x)$ is the generalized Riemann zeta-function
\cite{bat}. As in the previous section,
let us further introduce the dimensionless function
$V_2(x,y)$$=3\pi(eH)^{-3/2}V_{H_2}(\sigma,\Delta,\Delta^*)$:
\begin{eqnarray}
\label{eq.34}
V_2(x,y)=\frac{\tilde A x^2}2+\frac {\tilde B y^2}2+\frac 23|x|^3
+\sqrt{x^2+y^2}-2\zeta (-1/2, x^2+y^2),
\end{eqnarray}
where $x=\sigma/\sqrt{eH}$, $y=2|\Delta|/\sqrt{eH}$ and $\tilde
A,\tilde B$
are the same parameters, as in the relation (\ref{eq.200}).
Clearly, this expression differs from the corresponding quantity
(\ref{eq.200}) for a non-abelian background field.

The reader, who is not interested in following the details of our
investigation of a global minimum point for  $V_2(x,y)$, can at once
look at the phase portrait of the model in terms of $\tilde A,\tilde
B$.
Qualitatively it is the same as the phase portrait for the function
$V_1(x,y)$ (see Fig. 2), i.e. it contains only three different Phases
II, III and IV. Details of calculations are quoted in Appendix B.

\section{Chromomagnetic catalysis of color
superconductivity}

Let us now analyse in more details the phase portrait of the model
(1), this time in terms of $A,B,eH$. In particular, we shall
describe phase transitions which occur for arbitrary fixed $A,B$
and with varying values of $H$. In general, our discussions
concern both cases with vector-potentials of types I
and II simultaneously. However, where it is necessary, we indicate
to what type of vector-potential the information is related.
First of all a general remark:
If $A,B$ are fixed and $H$ is varied from 0 to $\infty$, then in
the plane $(\tilde A,\tilde B)$ one moves along some ray (which
depends on $A,B$) from infinity to the origin (this fact simply
follows from
the definition of $\tilde A,\tilde B$).

\underline{\bf The case $A,B>0$.} (In this case we have a weak
coupling
for both bare constants: $G_{1,2}<G_c\sim \pi^2/\Lambda$.) At $H=0$
this
choice of parameters
corresponds to the unbroken Phase I (see Fig.1). If $A,B$ are
fixed in such a way, that $A>B$ 
(in terms of bare coupling constants this means  
$\frac{1}{G_1}>\frac{1}{2G_2}+\frac{4\Lambda}{3\pi^2}$), 
then at $H\neq 0$ we
have in the
$(\tilde A,\tilde B)$ plane of Fig. 2 a ray which is located above
the line $\tilde l_1$, i.e. is 
in the Phase III. Hence, in this case the external
chromomagnetic field induces (catalyses) the dynamical generation
of a nonzero diquark condensate.
Here in the point $H_c=0_+$ one has a second order phase transition
from Phase I to the phase with color superconductivity.
At varying values of $H$ the diquark condensate
behaves, e.g. in the case with vector-potential II,
in the following way\footnote{In order to find
$<qq>\sim $$<\Delta>$ and $<\bar qq>\sim <\sigma>$, one should
multiply the coordinates of the global minimum point of
the functions $V_1(x,y)$ or $V_2(x,y)$ (see Appendixes A and B,
correspondingly) by the quantity $\sqrt{eH}$.
For example, in the case under consideration
 $<qq>\sim$$<\Delta>= y_0(\tilde B)\sqrt{eH}/2$, where the function
 $y_0(\tilde B)$ and some of its properties are presented in
formula (B.4) of Appendix B.}:
\begin{equation}
<qq>\sim feH~~~~\mbox{at}~~~H\to 0~,~~~<qq>\sim
\sqrt{eH}~~~~\mbox{at}~~~H\to\infty.
\end{equation}
(The chiral condensate in this case  is identically zero.)

If $A<B$ 
and the external chromomagnetic field $H$ varies
in the interval
$H\in (0,\infty)$, then points in the $(\tilde A,\tilde B)$ plane
vary along a
ray $r$ (see Fig. 2). If $H\to 0_+$, we are at the infinite end of
this ray, i.e.  in the Phase II of the model (see Fig. 2).
Hence, in the point $H_c=0_+$ the chromomagnetic field induces
a dynamical chiral symmetry breaking phase transition (a second
order phase transition), but a diquark condensate is not produced.
 These properties of the
vacuum are not changed for sufficiently small values of $H$ such
that $H<H_c(A,B)$.  The value $H=H_c(A,B)$ corresponds to the
point $o$ (see Fig. 2) in which the ray $r$ crosses the line $\tilde
l_2$ and passes to the $\Omega$-region where chiral and diquark
condensates are both nonvanishing.  So, with growing values of $H$ in
some point $H_c(A,B)$ one has a second order phase transition
from the Phase II to the mixed Phase IV. In Table 1 the results of a
numerical investigation of $H_c(A,B)$ as a function of
$A$ are presented for some fixed values of $B/A$ in the case
of a vector-potential of type I.
One can see from this table that if the ratio $B/A$ is fixed,
then $H_c(A,B)$ increases with $A$ as $A^2$. It is also clear
that for each fixed value of $A$ the quantity $H_c(A,B)$ is a
growing function of $B$.

The behaviour of condensates in the case $A<B$ and for a
vector-potential II
 are the following:
\begin{equation}
\label{eq.370}
<\bar qq>\sim geH~~\mbox{at}~~H\to 0,~~
<\bar qq>\sim <\sigma>\equiv \frac {(B-A)}2~~\mbox{at}~~H_c(A,B)\leq
H,
\end{equation}
\begin{equation}
<qq>\equiv 0~~\mbox{at}~~H\leq H_c(A,B),~~~~~
<qq>\sim\sqrt{eH}~~\mbox{at}~~H\to\infty.
\label{eq.380}
\end{equation}

So, at $H\neq 0$ the Phase I of the model is completely absent in
the phase structure of the model for both types of external
chromomagnetic fields.

\underline{\bf The case $A<0,2A<3B$.} In this case at $H=0$ one has
a Phase II of the theory (see Fig. 1) with spontaneously broken
chiral symmetry. If the external chromomagnetic field $H$ varies in
the
interval $H\in (0,\infty)$, then in the $(\tilde A,\tilde B)$ plane
there is a ray which crosses the line $\tilde l_2$ in some definite
point. If $H\to 0_+$, we are in the infinite end of this ray, i.e.
in the Phase II of the model. However, in contrast to the
previous case with $A,B>0$ and $A<B$, in the present case the
value $H=0_+$ is no more the point of a phase transition.  (At
$A<0$, when the bare coupling constant $G_1$ takes a supercritical
value $G_1>G_c$,
the origin of chiral symmetry breaking is the rather
strong supercritical quark-antiquark attraction, but not the
chromomagnetic field. In this case the external chromomagnetic field
only stabilizes the vacuum with chiral symmetry breaking \cite{1,3}.)
 If $H$ increases, we move along this ray
to the origin of the $(\tilde A,\tilde B)$-plane.  Hence, starting
from some value $H_c(A,B)$, we are in the region $\Omega$ (see
Fig. 2), where besides $<\bar q q>\neq 0$ the diquark condensate
is nonzero as well. So, at sufficiently high values of
$H>H_c(A,B)$ the Phase II of the theory is transformed into a
mixed Phase IV.

The influence of the vector-potential II on the chiral Phase II
of the model is realized in the following behaviour of condensates:
\begin{equation}
\label{eq.371}
<\bar qq>\sim <\sigma>=-\frac A6~~\mbox{at}~~H\to 0,~~
<\bar qq>\sim <\sigma>\equiv \frac {(B-A)}2~~\mbox{at}~~H_c(A,B)\leq
H,
\end{equation}
\begin{equation}
<qq>\equiv 0~~\mbox{at}~~H\leq H_c(A,B),~~~~~
<qq>\sim\sqrt{eH}~~\mbox{at}~~H\to\infty.
\label{eq.381}
\end{equation}

\underline{\bf The case $B<0, A>B$}. In this case at $H=0$ there is a
perfect (not
mixed) color superconducting Phase III of the theory (see
Fig. 1).  One can easily show that now for all values
of $H$ only the diquark condensate $<qq>$ is nonzero.  This
vacuum is chirally invariant, but the U$_{\rm em}(1)$
as well as color SU$_c(3)$ symmetries are broken down. It is possible
to show
that in this case
\begin{equation}
<qq>\sim <\Delta>= -B/4~~~~\mbox{at}~~~H\to 0~,~~~~~<qq>\sim
\sqrt{eH},~~~~\mbox{at}~~~H\to\infty,
\end{equation}
i.e. the external chromomagnetic field even enhances the color
superconductivity.

\underline{\bf The case $A<B, 2A>3B$.} Analyzing here the behavior of
the quark condensates in a similar way as in the previous cases, one
can easily establish that the vacuum properties are not changed
with growing values of $H$. Hence, at $H=0$ as well as at $H\neq 0$
there is a mixed Phase IV with  nonzero quark and diquark
condensates.
Remark that the action of an external chromomagnetic field on the
mixed phase
does not change the value of the chiral condensate: It is the same as
at $H=0$,
where $<\bar qq>\sim$$<\sigma>\equiv (B-A)/2$.
However, the diquark condensate depends on the value of $H$:
\begin{equation}
<qq>\sim <\Delta>=\sqrt{B^2/16-(B-A)^2/4}~~\mbox{at}~~H\to
0,~~~<qq>\sim \sqrt{eH},~~\mbox{at}~~H\to\infty
\end{equation}

In conclusion, let us remark that for arbitrary fixed
parameters $A,B$ and in the presence of sufficiently large
values of external chromomagnetic fields of both types there arises a
nonzero diquark condensate $<qq>\neq 0$, i.e. the color
superconducting
phase of the model is realized.
If $A>3B/2,~B<0$ (i.e. for sufficiently high values of $G_2>G_c$),
then $<qq>\neq 0$ even at $H=0$ (in this case the external
chromomagnetic field
enhances the CSC). However, for other
regions of the $(A,B)$-plane the nonzero external chromomagnetic
fields
catalyse the generation of $<qq>\neq 0$. The critical value of $H$,
at which color superconductivity is induced, may be $0_+$
(if $A,B>0,~A>B$), or some finite value $H_c(A,B)\neq 0$
(in the last case we have not a perfect, but mixed color
superconducting phase in which the diquark condensate
coexists with the chiral condensate $<\bar q q>\neq 0$).
Apart from this, in the presence of a chromomagnetic field,
the phase portrait of the model does not contain the symmetric phase.

\section{Summary and conclusions}

In the present paper the phase structure of a 
(2+1)-dimensional
four-fermionic NJL-type of model (1) with two coupling constants
was investigated admitting nonzero background vector-potentials of
two nonequivalent types I and II (see (\ref{eq.k2})-(\ref{eq.k1})).
In the framework of such a model the external
vector-potential might be thought to simulate such a nonperturbative
feature of the real QCD-vacuum like a nonzero gluon condensate $<FF>=H^2$. 
The structure of
the Lagrangian (1) permits us, in particular, to consider then the
competition between
chiral $<\bar qq>$ and diquark $<qq>$ condensates and to get some
insight into the role of the gluon condensate as a possible catalyst of
color superconductivity.

It is well-known that color-superconducting quark matter
with two quark flavors arises by the condensation of
color antitriplet diquark Cooper pairs.
 The condensate breaks the
SU$_c(3)$ symmetry down to SU$_c(2)$. Hence, the three gluon
fields  corresponding to the generators
of unbroken SU$_c(2)$ stay massless and the remaining five gluon 
fields receive a mass by the Higgs mechanism (Meissner effect). We have 
studied the influence
of the external chromomagnetic fields living in an unbroken SU$_c(2)$
subgroup of SU$_c(3)$, i.e. having a form
$H^a=(H^1,H^2,H^3,0,...,0)$,
on the formation of the color diquark condensate. Using a global
SU$_c(2)$
color rotations one can bring this field to the form
$H^a=(0,0,H,0,...,0)$
which corresponds to the above mentioned vector-potentials
(\ref{eq.k2})-(\ref{eq.k1})).

The main conclusion from our investigations is that at zero
chemical potential the external chromomagnetic fields of these type
are a good
catalyst of color superconductivity. (Earlier, it was shown that
external (chromo)magnetic fields catalyse dynamically the spontaneous
breaking of chiral symmetry in some (2+1)-dimensional four-fermionic
models \cite{2,krive,1,3}. It turns out that this is a particular
manifestation of the so-called magnetic catalysis effect (see e.g.
\cite{gus,dit,sem}), which has a rather universal model independent
character.)
 Indeed, we have shown that for sufficiently small bare
coupling constants $G_{1,2}<G_c$$\sim\pi^2/\Lambda$, i.e. for such
values of $G_{1,2}$
at which for $H=0$ one has a symmetric Phase I of the theory (see
Fig. 1), the pure CSC phase ($<\bar qq>=0$, $<qq>\neq 0$) is realized
in the model at infinitesimally small values of the external
chromomagnetic field
$H$ if $G_2>G_1$ (in terms of $A, B$ that means $A>B$). If $G_2<G_1$
($A<B$), then, first, a chirally breaking phase transition induced
at $H=0_+$ (chromomagnetic catalysis of chiral symmetry
breaking at which $<\bar qq>\neq 0$, $<qq>=0$) occurs. After that,
with
growing values of $H$, at some point $H=H_c$ there is a second phase
transition to the mixed phase of the theory,
where both condensates $<\bar qq>$ and $<qq>$ are nonzero (both
phase transitions are continuous second order ones).

The action of an external chromomagnetic field on the chiral Phase II
of the theory (see Fig. 1) is to induce $<qq>\neq 0$ at some
critical point $H=H_c\neq 0$, thus drastically changing the vacuum
properties and transposing the system into a mixed Phase IV.

Finally,
we should mention that the ground states of Phases III and IV are not
changed under the influence of an above mentioned external
chromomagnetic field. So,
the external chromomagnetic fields living in an unbroken SU$_c(2)$
subgroup of SU$_c(3)$ only enhances the CSC phenomenon.

Notice that all the above mentioned effects are observed in the
presence of both
vector-potentials I,II.

To our opinion there exists a deep connection
between the chromomagnetic catalysis of color superconductivity
and  chiral symmetry breaking, induced by external chromomagnetic
fields \cite{1}-\cite{3}. This assumption is based on the existence of
the Pauli--G\"ursey (PG) transformation \cite{pauli}, mixing quarks and
antiquarks, due to which some phenomena in the $\bar qq$-channel
can have its analogy in the $qq$-channel. In particularly, this suggests
that diquark condensation might be understood as the properly
PG-transformed chiral condensation. However, 
the detailed consideration of this question is not the subject of the present 
paper and will be investigated elsewhere.

Moreover, 
in the nearest future we are going to include into our consideration
of the simple NJL model (1) a nonzero chemical potential $\mu$  
in addition to the external chromomagnetic fields.
 Recently, in the framework of
NJL models \cite{vsh} the influence of $\mu$ and an external
magnetic field on the chiral properties of the vacuum were
considered.
Apart from discovering different kinds of magnetic oscillations
(relativistic van Alphen--de Haas effect) in the strongly
interacting quark systems,
there it was also found that in the NJL model at nonzero baryon
density the chiral
symmetry must be restored at sufficiently large values of the
magnetic field. By analogy with \cite{vsh}, we expect that the diquark
condensate
$<qq>$ should disappear 
in the case of nonzero baryon density
for a sufficiently strong external chromomagnetic
field, i.e. at large values of the gluon condensate.

\section*{Acknowlegments}

We wish to thank V. A. Miransky and Y. Nambu for interesting discussions and
A.K. Klimenko for numerical calculations.
One of us (D. E.) acknowledges the support provided to him by the
Ministry of Education of Japan (Monbusho) for his work at RCNP of Osaka
University.
This work is supported in part by the Russian Foundation of Basic
Researches (project 98-02-16690) as well as by DFG-Project 436 RUS
113/477/4.

\subsection*{\rightline{\underline{Appendix A}}}
\subsection*{Investigation of the global minimum point of $V_1(x,y)$}

It follows from (\ref{eq.22}) in the text that the function $F(z)$
monotonically
increases on the interval $z\in (0,\infty)$ and
$F(0_+)=-\infty$, $F(+\infty)=+\infty$. Hence, for arbitrary
fixed values of $\tilde A,\tilde B$ there exist only two real numbers
$x_0(\tilde A)>0,$ $y_0(\tilde B)>0$, such that the two pairs
$(x_0,0)$,
$(0,y_0)$ as well as the trivial one $(0,0)$ are solutions
for the system of stationarity equations (\ref{eq.21}).  (The
$x_0(\tilde A)$ and $y_0(\tilde B)$ are zeros of the functions which
are located inside the first (at $y=0$) and second (at $x=0$)
pair of braces in (\ref{eq.21}), respectively.) Furthermore, one
can easily see that $\partial V_1/\partial x<0$ if $y=0$ and $x\in
(0,x_0(\tilde A))$ as well as $\partial V_1/\partial y<0$ if $x=0$
and
$y\in
(0,y_0(\tilde B))$. This means that the quantities $V_1(x_0,0)$ and
$V_1(0,y_0)$ are smaller, than $V_1(0,0)$. So, for arbitrary
finite real values $\tilde A,\tilde B$ the global minimum point of
the function $V_1(x,y)$ cannot lie in $(0,0)$. Due to this reason
the symmetric phase is absent at all in the phase structure of the
model.

In the next formulae some properties of $x_0(\tilde A)$ and
$y_0(\tilde B)$
are presented:
$$
x_0(\tilde A)\cong 2e^{-\tilde A-2}~~~\mbox{at}~~\tilde
A\to+\infty,~~x_0(0)=0.147...,
~~x_0(\tilde A)\cong -\tilde A/6~~~\mbox{at}~~\tilde A\to-\infty,
\eqno(A.1)
$$
$$
y_0(\tilde B)\cong 2e^{-\tilde B-2}~~~\mbox{at}~~\tilde
B\to+\infty,~~y_0(0)=0.183...,
~~y_0(\tilde B)\cong -\tilde B/4~~~\mbox{at}~~\tilde B\to-\infty.
\eqno(A.2)
$$

It follows from (\ref{eq.21}) that there may exist (but not for
all values of $\tilde A,\tilde B$) one more solution 
$(x_1(\tilde A,\tilde B),y_1(\tilde A,\tilde B))$
of the stationarity equations, 
where $x_1>0,y_1>0$. (For this solution the functions which
are inside both braces in (\ref{eq.21}) take zero values.)
Evidently, we have
$$
x_1(\tilde A,\tilde B)=(\tilde B-\tilde A)/2,~y_1(\tilde A,\tilde
B)=\sqrt{y_0^2(\tilde
B)-
(\tilde B-\tilde A)^2/4}.
\eqno(A.3)
$$
From (A.3) one can easily see that this type of
solutions for equations (\ref{eq.21}) exists inside
the region $\Omega$ of the $(\tilde A,\tilde B)$-plane (see also Fig.
2):
$$
\Omega=\{(\tilde A,\tilde B)~:~\tilde B-2y_0(\tilde B)<\tilde
A<\tilde B\}.
\eqno(A.4)
$$
Using equations (\ref{eq.21}) one can find the following
values of the potential (\ref{eq.200}) in its stationary points:
$$
V_1(x_0(\tilde A),0)=\frac{\sqrt{1+x^2_0(\tilde A)}}6(1-2x^2_0(\tilde
A))-\frac23x^3_0(\tilde A),\eqno(A.5)
$$
$$
V_1(0,y_0(\tilde B))=\frac{\sqrt{1+y^2_0(\tilde B)}}6(1-2y^2_0(\tilde
B))-\frac13y^3_0(\tilde B),\eqno(A.6)
$$
$$
V_1(x_1(\tilde A,\tilde B),y_1(\tilde A,\tilde
B))=\frac{\sqrt{1+y^2_0(\tilde B)}}6
(1-2y^2_0(\tilde B))-\frac13y^3_0(\tilde B)
-\frac{(\tilde B-\tilde A)^3}{24}. \eqno(A.7)
$$
On the line $\tilde l_1=\{(\tilde A,\tilde B):\tilde A=\tilde B\}$,
which is a
part of the boundary for the region $\Omega$, we have
$(x_1,y_1)\equiv
(0,y_0)$.  Hence, on this line $V_1(0,y_0)\equiv$
$V_1(x_1,y_1)$. Comparing (A.6) and (A.7), we
see that inside the $\Omega$-region $V_1(0,y_0)>$$V_1(x_1,y_1)$.

The other part of boundary for the region $\Omega$ is the line
$\tilde
l_2=\{(\tilde A,\tilde B):\tilde A=\tilde B-2y_0(\tilde B)\}$.  With
the help
of the stationarity equations it is possible to show that on this
line
the following relations are also fulfilled: $\tilde B=\tilde
A+2x_0(\tilde A)$, $x_0(\tilde A)=y_0(\tilde B)$. As a consequence,
we
have $(x_1,y_1)\equiv (x_0,0)$ as well as $V_1(x_0,0)\equiv$
$V_1(x_1,y_1)$ on the line $\tilde l_2$.

Numerical investigations show that inside the region $\Omega$
there is a line on which $V_1(x_0,0)$$=V_1(0,y_0)$.
Further,  it is important to remark
that the derivative of the function $V_1(0,y_0(\tilde B))$ with
respect
to
$\tilde B$ as well as the corresponding partial derivative of
the function $V_1(x_1(\tilde A,\tilde
B),y_1(\tilde A,\tilde B))$  are positively defined quantities
in their
regions of definition.  Now, taking into account all the above
mentioned facts, it is possible to assert that in Fig. 2 the
phase portrait of the model (1) in the presence of a nonzero
vector-potential (\ref{eq.k2}) is presented in terms of $\tilde A$
and $\tilde B$. This means that above the line $\tilde l_1$ there is
the Phase III of the model, which corresponds to the $V_1(x,y)$
global minimum point of the form $(0,y_0)$. (In this case
$<\sigma>=0$, $<\phi>\neq 0$.) Below the curve $\tilde l_2$ the
effective potential global minimum has the form $(x_0,0)$. So in
this region the Phase II is located, since for such values of
$\tilde A, \tilde B$ the model has a vacuum with $<\sigma>\neq 0$,
$<\phi>=0$. Finally, inside the $\Omega$-domain there is a mixed
Phase IV, since here the global minimum point $(x_1,y_1)$
corresponds to the vacuum with both
nonzero condensates $<\bar q q>$ and $<qq>$.

\subsection*{\rightline{\underline{Appendix B}}}
\subsection*{Investigation of the global minimum point of
$V_2(x,y)$}

In this Appendix we present the search of the global minimum
point for
the potential $V_2(x,y)$
as well as its dependence on the parameters $\tilde A,\tilde B$.
Since this
function is symmetric under two discrete transformations $x\to -x$
and $y\to -y$, it is sufficient to study it only in the region
$x,y\geq 0$. The stationarity equations for $V_2(x,y)$ take
the form
$$
\partial V_2(x,y)/\partial x\equiv x\{\tilde
A+2x+(x^2+y^2)^{-1/2}-2\zeta (1/2,
x^2+y^2)\}=0,
\eqno(B.1)
$$
$$
\partial V_2(x,y)/\partial y\equiv y\{\tilde
B+(x^2+y^2)^{-1/2}-2\zeta
(1/2,
x^2+y^2)\}=0.
\eqno(B.2)
$$
One can see from (B.1)-(B.2) that the first
derivatives of $V_2$ do not exist in the point $(0,0)$. (This
means that if the point $(x,y)$ tends to the origin
along different ways, the resulting expressions for the partial
derivatives at the point $(0,0)$ do not
 coincide.) In contrast, the function $V_1(x,y)$ is
differentiable in the point $(0,0)$. So, we need a special
investigation of this point. Let us put $y=0$ in the equation
(B.1). Then, using properties of the $\zeta(\nu,x)$-function
\cite{bat}, it is easily seen that at $y=0$
and $x\to 0_+$ the partial derivative $\partial V_2/\partial x$ tends
to
(-1). Analogously, at $x=0$ and $y\to 0_+$ the derivative $\partial
V_2/\partial y$ (B.2) tends to (-1) as well. This fact means
that for arbitrary values of $\tilde A,\tilde B$ the point $(0,0)$
cannot
be a global minimum for the potential $V_2(x,y)$.
So, in contrast to the case with $H=0$, the ground state with
intact initial symmetry is no more possible in the model (1) at
$H\neq 0$.
Such a property of the effective potential is a characteristic
feature
for a phenomenon which is called
(chromo)magnetic catalysis of dynamical symmetry breaking.
According to this effect the external (chromo)magnetic field
promotes in a great extent the spontaneous breaking of initial
symmetry of the theory (for more details, see the last section of the
present paper).

Similar to the case with non-abelian vector-potential of the type I,
in the
present consideration it is possible  to show that for arbitrary
values of $\tilde A,\tilde B$ the stationarity equations
(B.1)-(B.2)  have two solutions of the form
$(x_0(\tilde A),0)$ and $(0,y_0(\tilde B))$, where
$$
x_0(\tilde A)\cong 1/\tilde A~~~\mbox{at}~~\tilde
A\to+\infty,~~~~~
~~x_0(\tilde A)\cong -\tilde A/6~~~\mbox{at}~~\tilde A\to-\infty,~~~
\eqno(B.3)
$$
$$
y_0(\tilde B)\cong 1/\tilde B~~~\mbox{at}~~\tilde
B\to+\infty,~~~~~
~~y_0(\tilde B)\cong -\tilde B/4~~~\mbox{at}~~\tilde B\to-\infty.~~~
\eqno(B.4)
$$
(Here and in the following discussions of the present Appendix we
use the same notations $x_0(\tilde A)$ and $y_0(\tilde B)$ for the
solutions of stationarity equations as in the previous Appendix.
But one should remember that these functions have quite
different numerical values, than similar functions had in Appendix
A.)

From (B.1)-(B.2) it follows that only
for $(\tilde A,\tilde B)\in\Omega$, where $\Omega$ is defined
formally in
(A.4), there is a solution of the form
$(x_1(\tilde A,\tilde B),y_1(\tilde A,\tilde B))$, where
$x_1>0,y_1>0$.
These functions are given in (A.3).
There are no other solutions of the stationarity equations.

Using numerical and analytical methods, it is now possible to
compare the values of the effective potential in its stationary
points and thus to find the global minimum point of $V_2$ as
well as its dependence on the parameters $\tilde A,\tilde B$ of the
theory.
We omit the details of this investigation and present only the
results, which can be formulated in form of a phase portrait of the
model
in terms of $\tilde A,\tilde B$.

It turns out, that Figure 2, which is the phase portrait
of the model for nonzero vector-potential I, formally may serve as a
phase portrait of the model for nonzero external gauge field of
the type II as well. In both cases the line $\tilde l_1$ separates
the color
superconducting Phase III from the mixed Phase IV. Further, in both
cases,
the Phase IV is separated from the chiral Phase II by the line
$\tilde
l_2$,
which has the same analytic definition through
$x_0(\tilde A)$ and $y_0(\tilde B)$ (see the figure caption to Fig.
2).
Moreover, the leading asymptotic behaviours of the $\tilde
l_2$-curves
are in both
cases identical (at $\tilde A\to\infty$ we have instead the line
$\tilde
l_1$,
at  $\tilde A\to -\infty$ it is the line $2\tilde A=3\tilde B$).
Of course, since for type I and type II vector-potentials the
functions
$x_0(\tilde A)$ and $y_0(\tilde B)$ obey different stationarity
equations,
the line $\tilde l_2$ of case I does not coincide with the line
$\tilde
l_2$
of case II. Finally, we should stress again that for both types
of nonzero vector-potentials I and II, the symmetric Phase I of the
theory, which is present in the phase structure of the model at
$H=0$ (see Fig. 1), is absent at all.

\newpage
\begin{center}
{\bf Figure captions}
\end{center}
\vspace*{0.5cm}
 
{\bf Fig.1} Phase portrait of the model at vanishing chromomagnetic
field $H=0$.
The boundaries $l_1,l_2$ are defined by
$l_1=\{(A,B):~A=B\}$,~$l_2=\{(A,B):~2A=3B\}.$

{\bf Fig.2} Phase portrait of the model in the presence of a nonzero
vector-potential I. The symmetric Phase I is absent at all. The
boundaries $\tilde l_1$ and $\tilde l_2$ of the region $\Omega$
 are defined according to
$\tilde l_1=\{(\tilde A,\tilde B):\tilde A=\tilde B\}$,
$\tilde l_2=\{(\tilde A,\tilde B):x_0(\tilde A)=y_0(\tilde B)\}$.
At $\tilde B\to +\infty$ ($\tilde B\to -\infty$) the curve $\tilde
l_2$ approaches
 asymptotically the line $\tilde l_1$ (the line $2\tilde A=3\tilde
 B$). The curve $\tilde
l_2$ intersects the axes $\tilde A$ and $\tilde B$ in the points
(-0.36) and
(0.3), respectively. For fixed $A$ and $B$ and varying $H$ one
moves in the
$(\tilde A,\tilde B)$-plane along some ray $r$. At $H=H_c(A,B)$ there
is a phase transition
from the Phase II to the mixed Phase IV,
 if the ray $r$ intersects the line $\tilde l_2$ in some point $o$
(see the detailed discussion in the Section 6).

\begin{table}
\caption{The $A$-dependence of $H_c(A,B)$ for some fixed
values of the ratio $B/A$.}
\begin{center}
\begin{tabular}{|p{25mm}|c|c|c|c|c|c|} \hline
 $B/A$ & 100 & 10 & 2 & 1.5 & 1.1 & 1.05  \\ \hline
$eH_c(A,B)/A^2 $ & 112925.63& 967.55& 15.804& 5.1038&
0.6593& 0.3508 \\ \hline
\end{tabular}
\end{center}
\end{table}

\begin{figure}
\begin{center}
\unitlength=1mm
\special{em:linewidth 0.4pt}
\begin{picture}(152.67,150.33)
\linethickness{0.4pt}
\linethickness{1.0pt}
\put(80.33,80.00){\line(1,0){70.33}}
\put(80.33,80.00){\line(0,1){60.67}}
\bezier{202}(80.33,80.00)(53.31,64.00)(26.33,44.00)
\bezier{202}(80.33,80.00)(70.33,50.00)(60.33,20.00)
\put(76.67,138.67){\makebox(0,0)[cc]{A}}
\put(147.67,84.67){\makebox(0,0)[cc]{B}}
\put(114.33,113.33){\makebox(0,0)[cc]{Phase I}}
\put(114.33,103.33){\makebox(0,0)[cc]{$<\bar qq>=0,<qq>=0$}}
\put(114.00,57.67){\makebox(0,0)[cc]{Phase II}}
\put(114.00,47.67){\makebox(0,0)[cc]{$<\bar qq>\not=0,<qq>=0$}}
\put(44.67,113.33){\makebox(0,0)[cc]{Phase III}}
\put(44.67,103.33){\makebox(0,0)[cc]{$<\bar qq>=0,<qq>\not=0$}}
\put(58.00,56.00){\makebox(0,0)[cc]{Phase IV}}
\put(50.00,48.00){\makebox(0,0)[cc]{$<\bar qq>\not=0,$}}
\put(42.00,40.00){\makebox(0,0)[cc]{$<qq>\not=0$}}
\put(39.00,57.33){\makebox(0,0)[cc]{$l_1$}}
\put(68.67,33.33){\makebox(0,0)[cc]{$l_2$}}
\end{picture}
\end{center}
\caption{}
\end{figure}

\newpage
\begin{figure}
\begin{center}
\unitlength=1mm
\special{em:linewidth 1.0pt}
\linethickness{0.4pt}
\begin{picture}(130.67,137.67)
\put(51.33,81.00){\vector(0,1){60.67}}
\put(35.67,94.00){\vector(1,0){95.00}}
\put(55.33,88.33){\makebox(0,0)[lc]{$\Omega$}}
\put(45.00,77.67){\makebox(0,0)[cc]{Phase IV}}
\put(128.00,98.00){\makebox(0,0)[cc]{$\tilde B$}}
\put(54.67,139.67){\makebox(0,0)[cc]{$\tilde A$}}
\put(92.33,71.00){\makebox(0,0)[cc]{Phase II}}
\put(35.67,113.33){\makebox(0,0)[cc]{Phase III}}
\put(86.33,129.33){\makebox(0,0)[cc]{$\tilde l_1$}}
\put(50.33,56.00){\makebox(0,0)[cc]{$\tilde l_2$}}
\put(118.33,119.33){\makebox(0,0)[cc]{r}}
\put(77.33,101.33){\makebox(0,0)[cc]{o}}
\linethickness{1.0pt}
\bezier{404}(101.33,126.33)(53.00,85.33)(42.67,49.00)
\bezier{202}(51.33,94.00)(69.33,109.00)(99.33,134.00)
\bezier{202}(51.33,94.00)(33.33,79.00)(27.33,74.00)
\linethickness{0.2pt}
\put(51.33,94.00){\line(3,1){70.00}}
\end{picture}
\end{center}
\caption{}
\end{figure}


\end{document}